\documentclass[showpacs,preprintnumbers,amsmath,amssymb,pra]{revtex4}
\usepackage{url}
\usepackage[utf8]{inputenc}
\usepackage{amsopn}
\usepackage{amsmath}
\usepackage{amssymb}
\usepackage{graphicx}
\usepackage{hyperref}
\usepackage{subfig}

\newcommand{\Tr}{\mathrm{Tr}}
\newcommand{\1}{{\rm 1\hspace{-0.9mm}l}}
\newcommand{\Id}{\1}
\newcommand{\ii}{\mathrm{i}}
\newcommand{\dd}{\mathrm{d}}
\newcommand{\ee}{\mathrm{e}}
\begin{document}

\title{Quantum control with spectral constraints}

\author{{\L}ukasz Pawela}
\email{lukasz.pawela@gmail.com}
\affiliation{Institute of Theoretical and Applied Informatics, Polish Academy
of Sciences, Ba{\l}tycka 5, 44-100 Gliwice, Poland}

\author{Zbigniew Pucha{\l}a}
\email{z.puchala@iitis.pl}
\affiliation{Institute of Theoretical and Applied Informatics, Polish Academy
of Sciences, Ba{\l}tycka 5, 44-100 Gliwice, Poland}

\begin{abstract}
Various constraints concerning control fields can be imposed in the realistic
implementations of quantum control systems. One of the most important is the
restriction on the frequency spectrum of acceptable control parameters.  It is
important to  consider the limitations of experimental equipment when trying to
find appropriate control parameters. Therefore, in this paper we present a
general method of obtaining a piecewise-constant controls, which are robust with
respect to spectral constraints. We consider here a Heisenberg spin chain,
however the method can be applied to a system with more general interactions. To
model experimental restrictions we apply an ideal low-pass filter to numerically
obtained control pulses. The usage of the proposed method has negligible impact
on the control quality as opposed to the standard approach, which does not take
into account spectral limitations.
\end{abstract}

\date{30/IV/2012}



\maketitle

\section{Introduction}
One of the fundamental issues of the quantum information science is the ability
to manipulate the dynamics of a given complex quantum system. Since the
beginning of quantum mechanics, controlling a quantum system has been an
implicit goal of quantum physics, chemistry and implementations of quantum
information processing. 

If a given quantum system is controllable, i.e. it is possible to drive it into
a previously fixed state, it is desirable to develop a control strategy to
accomplish the required control task.  In the case of finite dimensional quantum
systems the criteria for  controllability can be expressed in terms of
Lie-algebraic concepts
\cite{albertini2002lie,d2008introduction,elliott2009bilinear}. These concepts
provide a mathematical tool, in the case of closed quantum systems, i.e. systems
without external influences. 

It is an important question whether the system is controllable when the control 
is performed only on a subsystem. This kind of approach is called a
\emph{local-controllability} and can be considered only in the case when the
subsystems of  a given system interact. As examples may serve coupled spin
chains or spin networks 
\cite{d2008introduction,burgarth2007full,burgarth2009local,puchala2012local}. 
Local-control has a practical importance in proposed quantum computer
architectures, as its implementation is simpler and the effect of decoherence is
reduced by decreased  number of control
actuators~\cite{PhysRevLett.99.170501,PhysRevB.81.085328}.

A widely used method for manipulating a quantum system is a coherent control
strategy, where the manipulation of the quantum states is achieved by applying
semi-classical potentials in a fashion that preserves quantum coherence. In the
case when a system is controllable it is a point of interest what actions must
be performed to control a system most efficiently, bearing in mind limitations 
imposed by practical restrictions. Various constraints concerning control fields
can be imposed in the realistic implementations of quantum control systems. One
of the most important is the restriction on the frequency spectrum of acceptable
control parameters. Such restrictions come in to play, for example, in an
experimental set up which utilizes an external magnetic
field~\cite{chaudhury2007quantum}. In the case of such systems, due to various
limitations, the application of piecewise-constant controls is not accurate. The
real realization of controls is somehow smoothed by some filter induced by an
experimental limitations. Thus, it is reasonable to seek control parameters in
the domain imposed by the experimental restrictions. 

In article~\cite{heule2010local} there has been discussed how the low-pass
filtering, i.e. eliminating high-frequency components in a Fourier spectra, on a
numerically obtained optimal control pulses affects a quality of performed
control. This approach makes a contact with experimental realizations, since it
implements the limitations or real quantum control systems.

In this paper we present a general method of obtaining a piecewise-constant
controls, which is robust with respect to low-pass filtering. The above means,
that elimination of high-frequencies in a Fourier spectra reduces the fidelity
only by a small amount. We utilize this approach to obtain numerically control
pulses on a Heisenberg spin chain
\cite{heule2010local,heule_controlling_2011,miszczak2011qubit}, however it can
be applied to a quantum system with more general interactions.

This paper is organized as follows.
In Section~\ref{sec:model} we provide a general description of a quantum mechanical control system.
In Section~\ref{sec:setup} we provide the description of the simulation setup used to test our model.
Section~\ref{sec:results} contains results obtained from numerical simulations and their discussion.
In Section~\ref{sec:conclusions} we provide a summary of the presented work and give some concluding remarks.
\section{Our model}\label{sec:model}
To demonstrate a method of obtaining piecewise-constant controls, which are
robust with respect to low-pass filtering we will consider an isotropic
Heisenberg spin-$1/2$ chain of a finite length $N$. The control will be
performed on the first spin only. The total Hamiltonian of the aforementioned
quantum control system is given by
	\begin{equation}
		H(t) = H_0 + H_c(t),
	\end{equation} 
where 
	\begin{equation}
		H_0 = J \sum_{i = 1}^{N - 1} 
		S_x^iS_x^{i+1}+S_y^iS_y^{i+1}+S_z^iS_z^{i+1},
	\end{equation}
is a drift part given by the Heisenberg Hamiltonian. The control is performed
only on a first spin and is Zeeman-like, i.e.
	\begin{equation}
		H_c(t) = h_x(t)S_x^1 + h_y(t)S_y^1.
	\end{equation}
In the above $S_k^i$ denotes $k^{\text{th}}$ Pauli matrix acting on the spin $i$.
Time dependent control parameters $h_x(t)$ and $h_y(t)$ are chosen to be
piecewise constant. Furthermore, as opposed to~\cite{heule2010local}, we do not
restrict the control fields to be alternating with $x$ and $y$, i.e. they can be
applied simultaneously (see e.g.~\cite{khaneja_optimal_2005} for similar
approach). For notational convenience, we set $\hbar = 1$ and after this
rescaling frequencies and control-field amplitudes can be expressed in units of
the coupling strength $J$, and on the other hand all times can be expressed in
units of $1/J$~\cite{heule2010local}.

The system described above is operator controllable, as it was shown
in~\cite{burgarth2009local} and follows from a controllability condition using a
graph infection property introduced in the same article. The controllability of
the described system can be also deduced from a more general condition utilizing
the notion of hypergraphs~\cite{puchala2012local}.

Since the interest here is focused on operator control, a quality of a control
will be measured with the use of gate fidelity,
\begin{equation}
	F = \frac{1}{2^N} |\Tr( U_T^\dagger U(h) )|,
\end{equation}
where $U_T$ is the target quantum operation and $U(h)$ is an operation 
achieved by control parameters $h$.

To obtain piecewise-constant controls, which are robust with respect to low-pass
filtering we will minimize the power in the high frequency part of a controls
Fourier spectrum. We will do so by minimizing the following functional 
\begin{equation}
	G = (1 - \mu) P  - \mu F,\label{eq:functional}
\end{equation}
where $F$ is the gate fidelity described above, $\mu$ is a weight assigned to fidelity
and $P$ is a contribution of high
frequencies in the total power of the control parameters. The above can be stated
as, if $y$ is a vector of Fourier coefficients for a control parameters $h$ of
length $n$, i.e.
\begin{equation}
y = Q h, \ \text{ where } \ Q = n^{-1/2}\{ \ee^{2 \pi \ii kl /n }\}_{k,l=0}^{n-1},\label{eq:qft}
\end{equation}
then $P$ is given by
\begin{equation}
P = \frac{\sum_{i=\frac{n}{2} - \Delta}^{\frac{n}{2}+\Delta}|y_i|^2}{|y|^2},\label{eq:power}
\end{equation}
for some $\Delta$ related to the cutoff frequency of low-pass filter.
\section{Simulation setup}\label{sec:setup}
To demonstrate the beneficialness of our approach, we study three- and four-qubit spin chains.
The control field is applied to the first qubit only. Our target gates are:
	\begin{equation}
		NOT_N = \Id^{\otimes N-1}\otimes \sigma_x,
	\end{equation}
the negation of the last qubit of the chain, and
	\begin{equation}
		SWAP_N = \Id^{\otimes N-2}\otimes SWAP,
	\end{equation}
swapping the states between the last two qubits.

For each of these cases we find two sets of control parameters. One with frequency constraints
and one without. Next, we calculate an appropriate filter and using these 
filtered values of control parameters, we calculate the fidelity of the 
quantum  operation. In each case 120 independent sets if control parameters were found.

We set the duration of the control pulse to $\Delta t = 0.2$ and the total number of pulses in each direction to 
$n=128$ for the three-qubit chain and $n=512$ in the four-qubit case. The weight of fidelity in 
equation~\eqref{eq:functional} is set to $\mu=1$ in the unconstrained case and to $\mu=0.05$ in the constrained case. 
Although the weight of the fidelity is small, the optimization still yields high fidelity values while maintaining low 
contribution of high frequencies in the power spectrum. We set the cutoff frequency in 
equation~\eqref{eq:power} to $\Delta=\frac{n}{4}$.

The applied filter is a frequency filter with the cutoff frequency equal to the frequency discriminated by the 
functional \eqref{eq:functional}. We consider this filter to be an ideal low-pass filter which leaves only the 
frequencies only in the interval $[-\omega_0,\omega_0]:\;f(\omega) = \Theta(\omega+\omega_0) - \Theta(\omega - 
\omega_0)$. We obtain the following expressions for the filtered control parameters \cite{heule_controlling_2011}:
\begin{equation}
	h_k = \frac{1}{\pi} \sum_{i = 1}^n h_{k,i}[a_{i+1}(t) - a_i(t)]
\end{equation}
where $k\in \{x,y\}$ and 
\begin{eqnarray}
	a_n(t) & = & \mathrm{Si}\left[\omega_0(nT -t)\right]\\
	\mathrm{Si}(x) & = & \int_{0}^{x}(\sin t / t)\dd t
\end{eqnarray}

We conduct our simulations using the BFGS gradient method. The calculation of
the gradient of the fidelity function  can be found in the work by Machnes
\emph{et al.} \cite{machnes_comparing_2011}

To calculate the gradient of $P$ one should note, an elementary fact concerning
differentiation of vector valued functions.  For a real vector $h$, we
define $y = A h$ for some fixed matrix $A$. Straightforward
calculations gives us
\begin{equation}
\frac{\partial |y_k|^2}{\partial h_l} = 2 \Re (\overline{A}_{kl} y_k).
\end{equation}
In the case when matrix is a quantum Fourier transform gate $A = Q$ defined in Eqn.~\eqref{eq:qft},
we obtain that
\begin{equation}
\frac{\partial |y_k|^2}{\partial h_l} = 2 \Re (\overline{Q}_{kl} y_k) = \frac{2}{\sqrt{n}} \Re (\ee^{-2 \pi \ii kl /n} 
y_k).
\end{equation}
This calculation is used to find the gradient of the contribution of high frequencies in the total
power given by Eqn.~\eqref{eq:power}.
\section{Results}\label{sec:results}
\begin{figure}[!h]
	\centering
	\subfloat[Unconstrained case]{\label{fig:not3}\includegraphics[width=9cm]{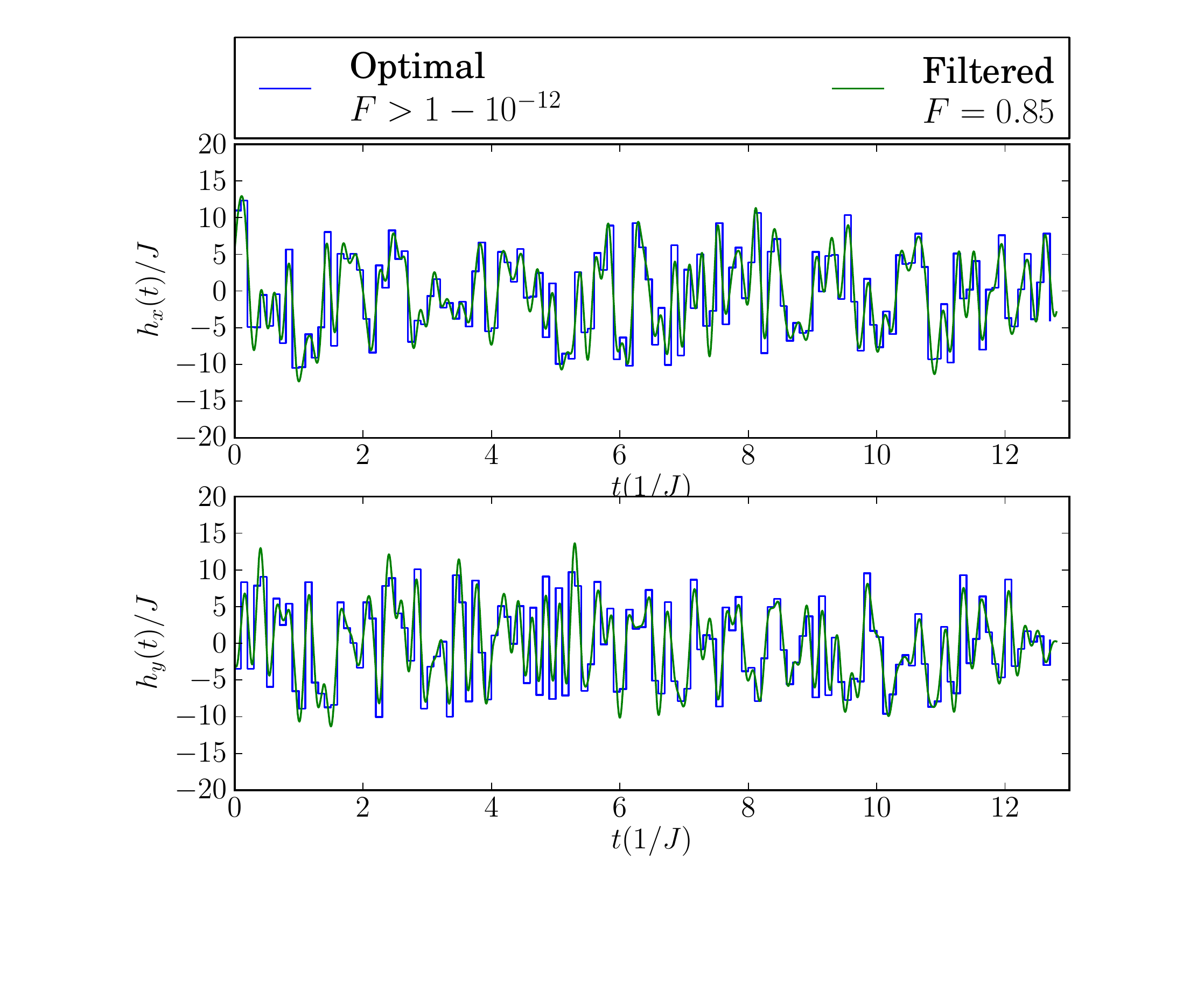}}
	\subfloat[Constrained case]{\label{fig:not3_fft}\includegraphics[width=9cm]{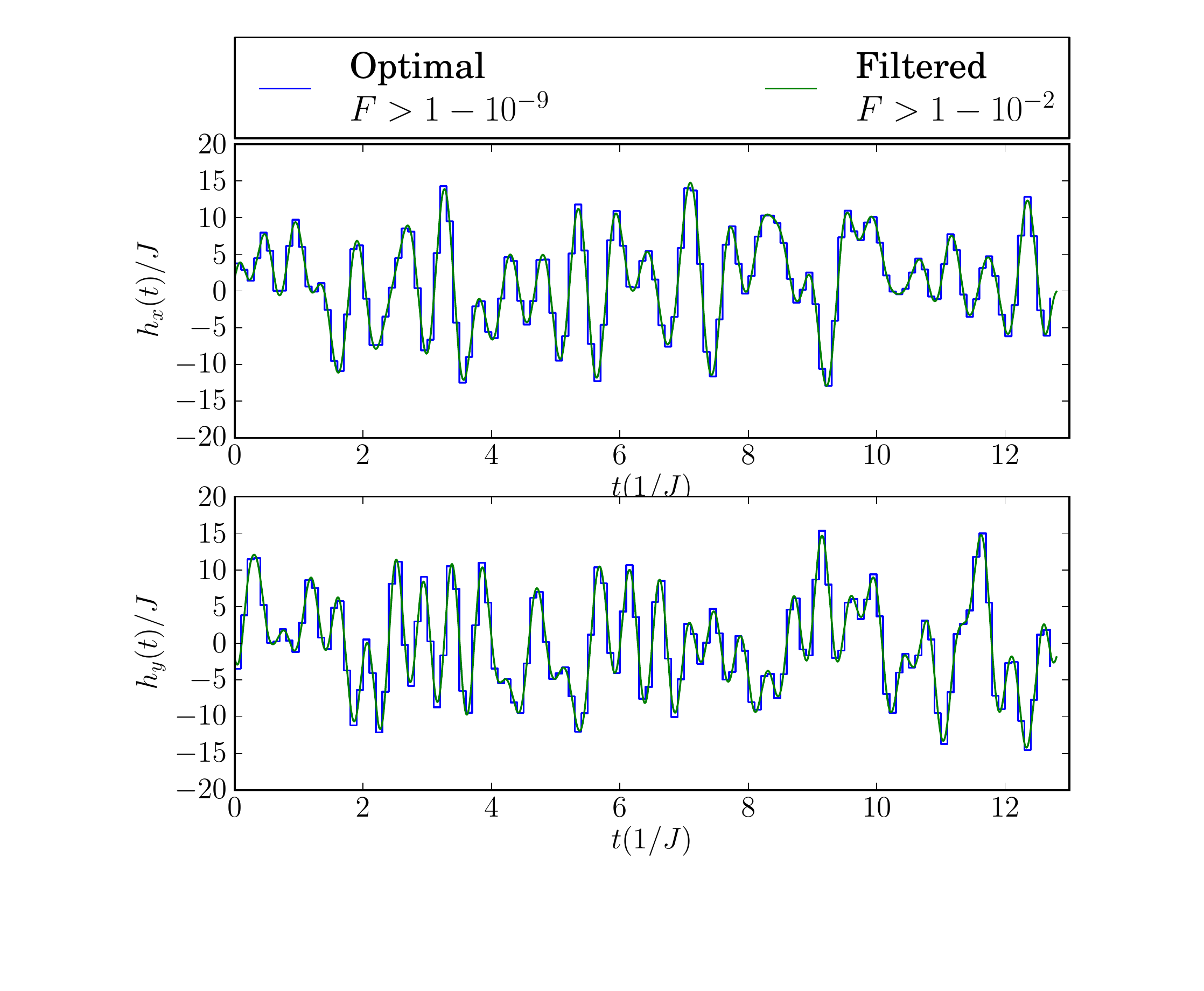}}
	\caption{$x$ and $y$ components of the control field for target gate $U_T = NOT_3$}
\end{figure}

Figure \ref{fig:not3} shows a plot of the control parameters for the target gate $U_T=NOT_3$ before and after 
applying the frequency filter. These parameters were found using with the value of the weight $\mu = 1$, resulting in 
no penalty for high frequency terms. Clearly, the signal after filtering differs from the original values. This is 
reflected in the values of the fidelity of the operation. Before filtering the fidelity is $F>1-10^{-12}$, however 
after filtering the value drops to $F=0.85$.

Next, in Figure \ref{fig:not3_fft} we show the plot of the control parameters for the target gate $U_T=NOT_3$, before 
and after applying the frequency filter. Only this time, the controls were found using the value of the weight $\mu = 
0.05$ resulting in a penalty for high frequency terms. A short glance reveals that the filtered parameters are almost 
the same as the original ones. This is reflected by the fidelity of the operation. Before filtering the fidelity is $F 
> 1-10^{-9}$ and after filtering it drops only to $F>1-10^{-2}$, which is still a satisfactory value. Hence, these 
sets of control parameters are well-suited for use in computations.

\begin{figure}[!h]
	\centering
	\subfloat[Unconstrained case]{\label{fig:swap3}\includegraphics[width=9cm]{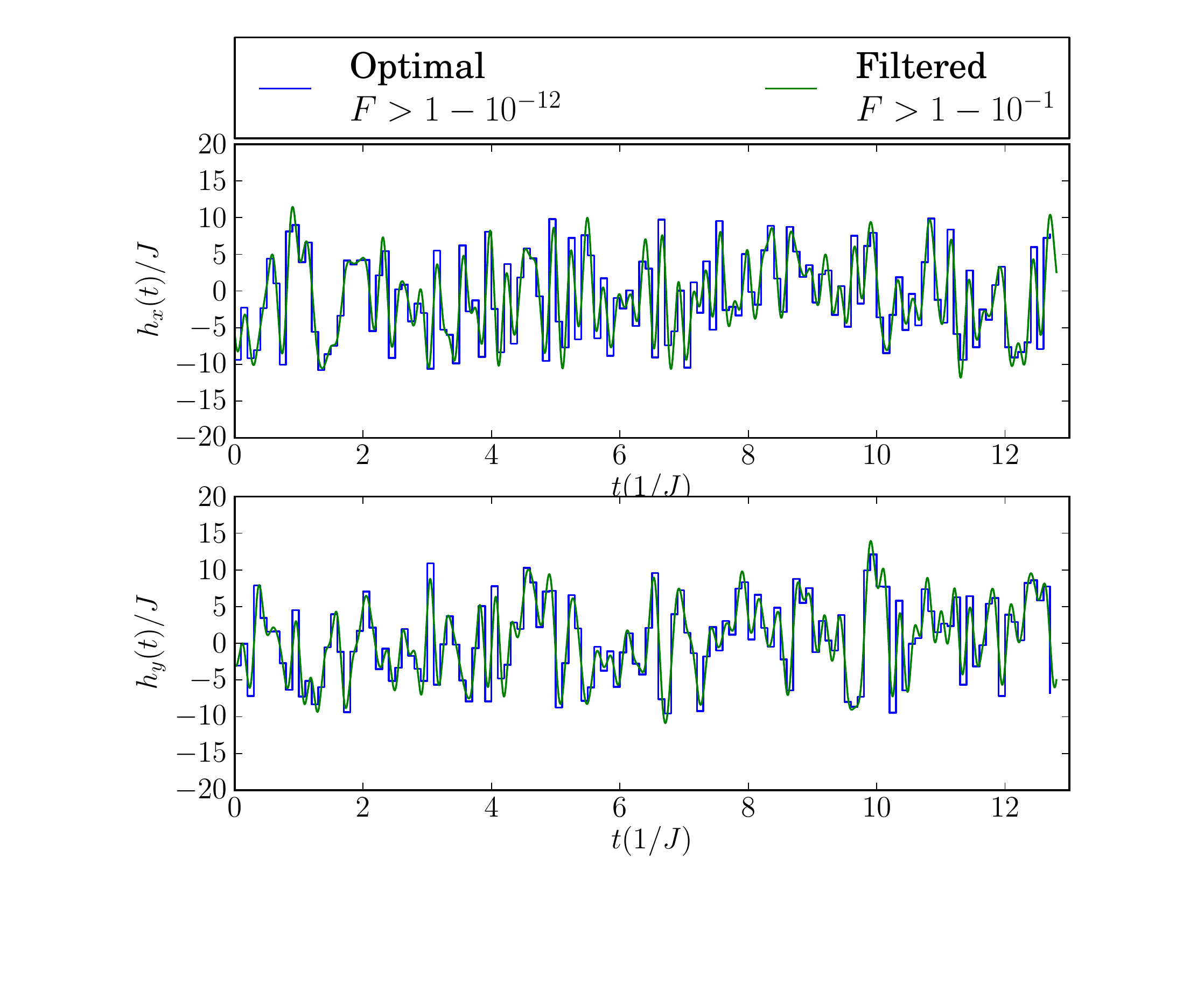}}
	\subfloat[Constrained case]{\label{fig:swap3_fft}\includegraphics[width=9cm]{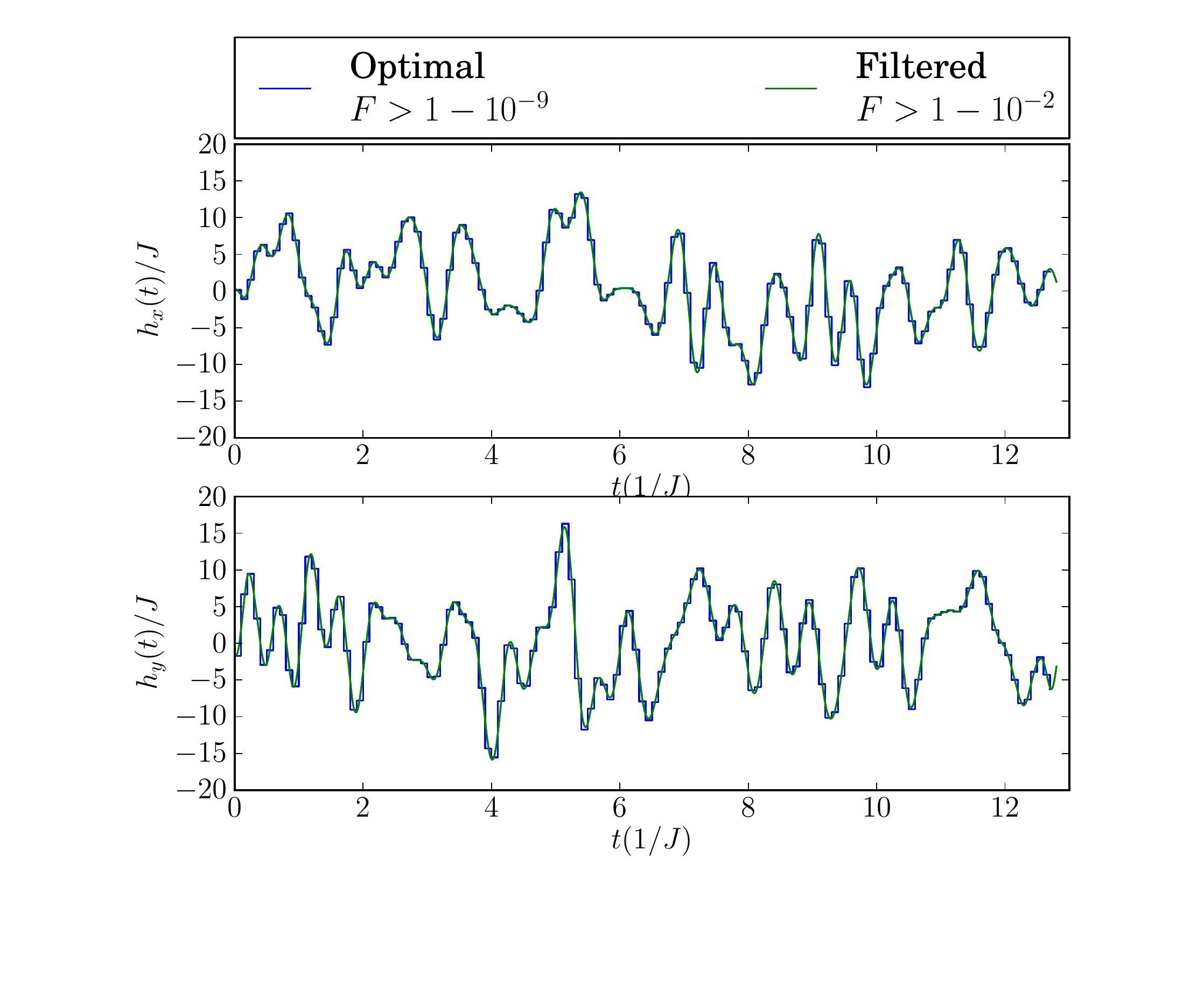}}
	\caption{$x$ and $y$ components of the control field for target gate $U_T = SWAP_3$}
\end{figure}

Figures \ref{fig:swap3} and \ref{fig:swap3_fft} show analogical results for the target gate $U_T=SWAP_3$. The 
qualitative results in this case are the same as for the $NOT_3$ target gate discussed earlier. Fidelity values before 
and after filtering are the same order as for the $NOT_3$ gate. Again, we reach a conclusion that these 
sets of control parameters are well-suited for use in computations.

Results obtained for the four-qubit chain (not shown here) qualitatively resemble the results for the three-qubit 
chain. In this case a penalty for high frequency terms of the control parameters, also leads to higher fidelities 
after filtering.

\begin{figure}[!h]
	\centering\includegraphics[width=9cm]{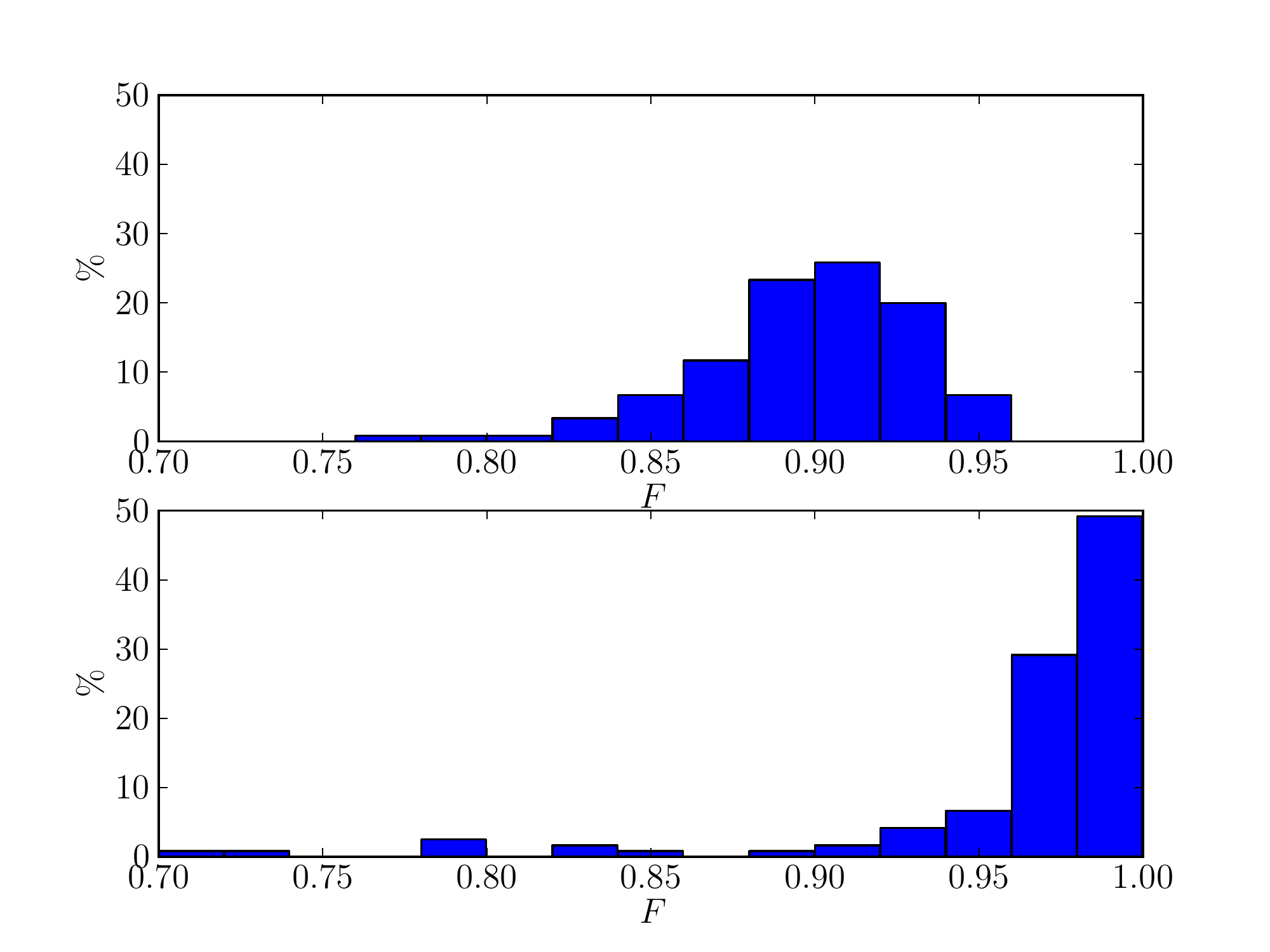}
	\caption{A histogram of the fidelities after filtering. Top: without considering spectral constraints, bottom: 
	with spectral constraints considered. Target gate $U_T =NOT_3$.} \label{fig:not3_hist}
\end{figure}

Figure \ref{fig:not3_hist} depicts the histograms of fidelities of filtered control parameters for the $NOT_3$ gate. 
The top plot shows the histograms for control parameters found without spectral constraints and the bottom one shows 
what happens when one takes spectral constraints into consideration. Clearly, a typical set of control parameters has 
a higher fidelity of operation when one considers spectral constraints in the optimization phase. Strictly speaking, 
around 80\% of the control parameters sets have a fidelity greater than $0.96$, whereas in the unconstrained case 
there are no control parameters sets that have such high fidelities. These facts lead to a conclusion that 
optimization with spectral constraints will lead to high fidelity of experimental realizations of the $NOT_3$ gate.

\begin{figure}[!h]
	\centering\includegraphics[width=9cm]{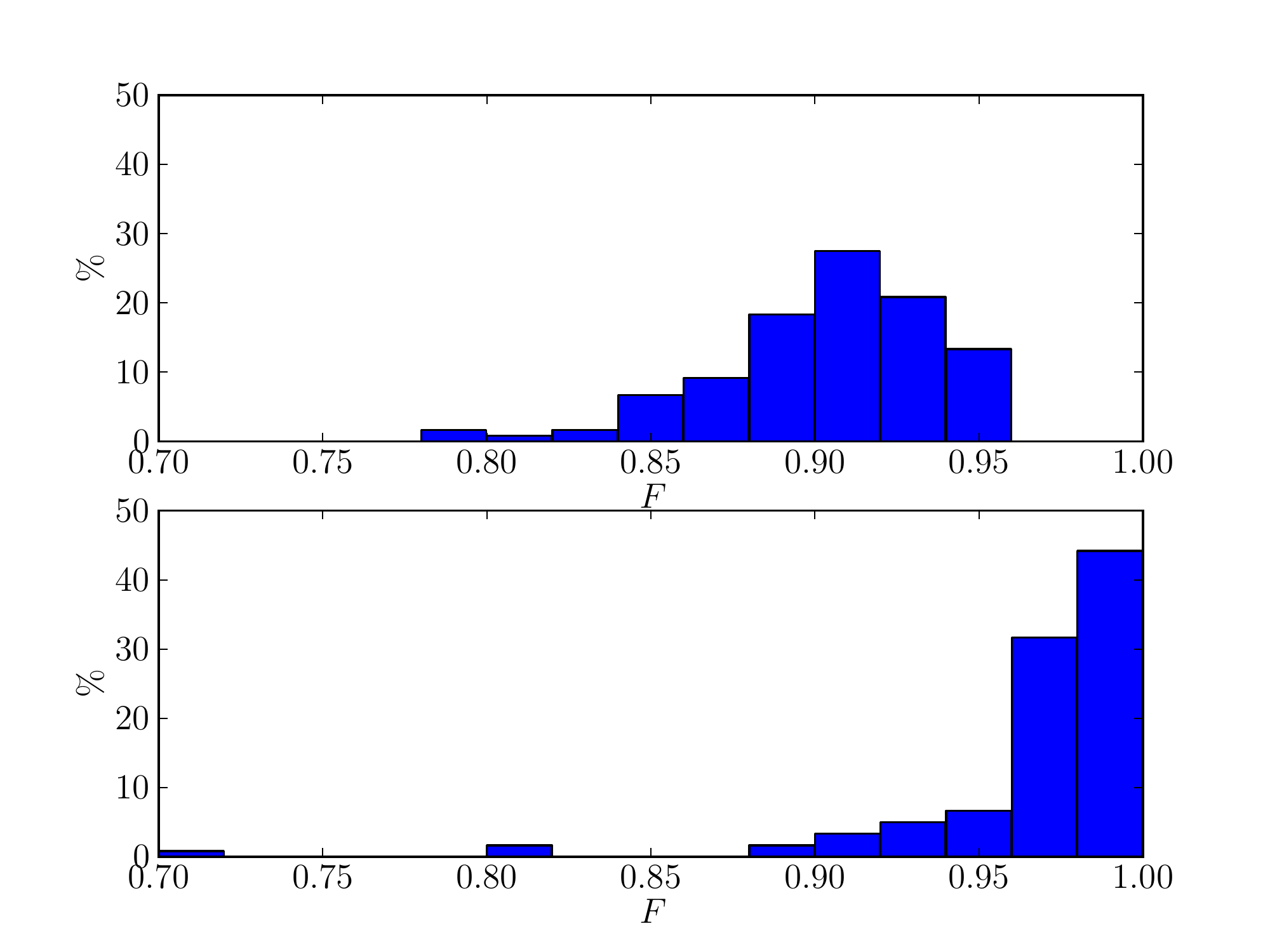}
	\caption{A histogram of the fidelities after filtering. Top: without considering spectral constraints, bottom: 
	with spectral constraints considered. Target gate $U_T = SWAP_3$.} \label{fig:swap3_hist}
\end{figure}

Analogical results for the $SWAP_3$ are shown in Figure \ref{fig:swap3_hist}. The top plot shows the histograms for 
control parameters found without spectral constraints and the bottom one shows what happens when one takes spectral 
constraints into consideration. In this case, around 75\% of all constrained control parameter sets have a fidelity 
higher than $0.96$ after filtering. Also, there are no unconstrained control parameter sets with fidelities in this 
range. Hence, spectral constraints imposed during the optimization step have lead to control parameters far less 
sensitive to experimental equipment limitations.
\section{Conclusions}\label{sec:conclusions}
We investigated the impact of spectral constraints imposed on the control parameters of a quantum operation on the 
fidelity of the quantum operation which they implement. In order to compare our approach with the unconstrained case, 
we apply an ideal low-pass to the control parameters.

We have shown that imposing spectral constraints on the control parameters leads to higher average fidelity of the 
quantum operation after appropriate filtering, than in the unconstrained case. These results are independent of the 
type of quantum operation and the number of qubits in the system under consideration.

Furthermore, the  requirement for smooth control parameters does not result in the increase of time necessary to 
conduct a quantum operation. Comparing with other research in the field \cite{heule_controlling_2011}, our times are 
on the same order.

Further work on this subject might take into account more subtle parameters of the experimental setup than the 
frequency cutoff of signal sources. For instance, one could wish to find control parameters that are far from 
transient characteristics of the experimental setup. This may lead to an enhancement of fidelities of operations 
achieved experimentally by eliminating unwanted signal roughness.

\begin{acknowledgements}
Work by {\L}.~Pawela and Z. Pucha{\l}a was supported by the Polish
National Science Centre under the grant number N N514 513340.
Numerical simulations presented in this work were performed on the
``Leming'' and ``{\'S}wistak'' computing systems of The Institute of
Theoretical and Applied Informatics, Polish Academy of Sciences.
\end{acknowledgements}

\bibliography{spectral_control}
\bibliographystyle{apsrev}

\end{document}